\documentclass[usenatbib,useAMS]{mn2e}
\usepackage{graphicx}
\usepackage{times}
\usepackage{aas_macros}
\title[]{Accretion-induced luminosity spreads in young clusters: evidence from stellar rotation}

\author[]{S.\,P.\ Littlefair$^{1}$, Tim Naylor$^{2}$, N.\,J.\, Mayne$^{2}$, Eric Saunders$^{3}$ \&
R.\,D\ Jeffries$^{4}$ \\
$^1$Dept of Physics and Astronomy, University of Sheffield, S3 7RH, UK \\
$^2$School of Physics, University of Exeter, Exeter, EX4 4QL, UK\\
$^3$Las Cumbres Observatory, 6740 Cortona Dr, Suite 102, Santa Barbara, CA 93117, USA\\
$^4$School of Chemistry and Physics, Keele University, Keele, Staffordshire, ST5 5BG, UK\\}

\date{\center{\Large Submitted for publication in the Monthly Notices of the
Royal Astronomical Society \\ 
\vspace{.5cm} \today}} 

\begin{document}
\maketitle

\begin{abstract} 
  We present an analysis of the rotation of young stars in the
  associations Cepheus OB3b, NGC 2264, NGC 2362 and the Orion Nebula
  Cluster (ONC). We discover a correlation between rotation rate and
  position in a colour-magnitude diagram (CMD) such that stars which
  lie above an empirically determined median pre-main sequence rotate
  more rapidly than stars which lie below this sequence. The same
  correlation is seen, with a high degree of statistical significance,
  in each association studied here. If position within the CMD is
  interpreted as being due to genuine age spreads within a cluster,
  then the stars above the median pre-main sequence would be the
  youngest stars. This would in turn imply that the most rapidly
  rotating stars in an association are the youngest, and hence those
  with the largest moments of inertia and highest likelihood of
  ongoing accretion. Such a result does not fit naturally into the
  existing picture of angular momentum evolution in young stars, where
  the stars are braked effectively by their accretion discs until the
  disc disperses.

  Instead, we argue that, for a given association of young stars,
  position within the CMD is not primarily a function of age, but of
  accretion history. We show that this hypothesis could explain the
  correlation we observe between rotation rate and position within the
  CMD.
\end{abstract} 


\section{Introduction}
\label{sec:introduction}
The rate at which star formation occurs is still an area of ongoing
debate. The two competing paradigms of rapid star formation on
dynamical timescales \citep[e.g][]{elmegreen00,hartmann01}, and slow
star-formation on timescales of several Myr \citep[e.g][]{shu87}, are
difficult to choose between on observational grounds. One argument in
favour of slow star-formation (SSF) has been the observation of
apparent age spreads in young clusters and associations
\citep[e.g][]{herbst82,sung98,pozzo03,dolan01}. These apparent age
spreads manifest themselves as a spread in the location of young stars
within a Hertzsprung-Russell (H-R) or a colour-magnitude diagram
(CMD); at a given luminosity stars with a range of
effective temperatures are observed, implying a range of radii or a spread in
ages. \cite{palla00} have used ages obtained from the H-R diagram to
argue that star formation takes place over $\sim$10\,Myr and
accelerates towards the present day. However, the reality of these
apparent age spreads has been questioned by various
authors. \cite{hartmann01} suggested a number of factors which could
produce a spread in the CMD or H-R diagram without requiring a genuine
spread in ages. These included photometric variability, inadequate
correction for variable extinction and the presence of unresolved
binaries. Since then, \cite{burningham05} have shown that photometric
variability cannot explain the observed spreads; nor can carefully
controlling for the effects of binarity, variable reddening and
contaminating light from the accretion disc explain the observed
spreads in the Orion Nebular Cluster \citep{dario10a} or LH 95 in the
Large Magellanic Cloud \citep{dario10b}. In addition \cite{jeffries07}
used a novel geometrical technique to show that the apparent age
spreads in the H-R diagram of the ONC are associated with a genuine
spread in stellar radii. These recent results have added support to
the idea that the luminosity spreads observed in young clusters and
associations are a genuine phenomenon, and not an observational
artifact. However, as accretion can affect the evolution of the
central star, inducing a spread of luminosities and radii in a
population which is co-eval, such spreads may not be attributable to a
spread in ages. Most studies agree that current accretion rates
typically observed in pre-main-sequence objects cannot cause the
significant spreads in luminosity (or radius) observed in stellar
populations \citep[e.g.][]{tout99,hartmann97,siess99}. However,
accretion and infall rates during the early, `assembly' phase of star
formation are expected to be much larger than those observed for more
evolved stars.

\cite{baraffe09} demonstrate that assembly phase accretion can significantly affect the radius
of the central star, causing accelerated contraction. Since the star's
radius changes on the thermal timescale, which also governs the
contraction of the star towards the main sequence, a star's radius
will carry the imprint of this elevated protostellar accretion for a
large fraction of its pre-main-sequence lifetime. \cite{baraffe09}
find that apparent age spreads of up to 10 Myr can be produced by
accretion rates of $\sim$10$^{-4}$M$_{\odot}$\,yr$^{-1}$. Such extreme
accretion rates would cause young stars to be much brighter than observed.
However, \cite{baraffe09} show that {\em episodic} accretion events with 
similar rates can still reproduce the observed luminosity spreads, without
causing a luminosity problem. There is growing evidence that accretion onto young stars shows large variations 
in accretion rate \citep[e.g.][]{enoch09}. Episodic accretion in not a 
prerequisite for significant luminosity spreads, but it can simultaneously
explain the observed luminosity spreads without causing anomolously high
luminosities and is more consistent with our current understanding of star formation and disc evolution
\citep{vorobyov06,vorobyov10,zhu09}.

Obtaining observational evidence for accretion-induced radius spreads is
not a straightforward task, as the accretion history of a young star
is not easily determined. One property which may be related to
accretion history is the stellar rotation rate.  Young stars typically
rotate with periods in the range of 1--10 days, which is a fraction of
their breakup speed. Whilst the underlying physical mechanism is not
yet clear \citep[see][for a discussion]{matt10}, there is strong
observational evidence that the population of stars with discs rotate
more slowly than those without disks
\citep[e.g.][]{edwards93,herbst00,littlefair05,rebull06,cieza07}. The
modelling work of \cite{rebull04a} showed that the period distribution
of young stars can be explained if accretion from a disc results in a
constant stellar spin period of around 7 days. Thus, the currently
accepted view of spin evolution of young stars is that stars which are
accreting are braked by the star-disc interaction and rotate
slowly. Once the disc disperses, or the accretion rate drops below
some critical threshold, the star will spin up again as it contracts
towards the main-sequence. This spin-up is slow; it is a direct result
of the contraction of the star and thus proceeds on the thermal
timescale. Therefore, the present day rotation rate of the star
depends upon the accretion history, although the exact relationship
will depend on the details of the braking mechanism and be complicated
by additional factors such as the stellar magnetic field strength.

If both luminosity and rotation rates are functions of the accretion
history, it is possible that a correlation between rotation and
luminosity exists. Motivated by this argument, we present an analysis
of the link between rotation rate and luminosity for four star forming
regions; the ONC, NGC 2264, Cep OB3b and NGC 2362.

\section{Results}
\label{sec:results}

The ONC, NGC 2264, Cep OB3b and NGC 2362 are all young ($\le 5$\, Myr)
star forming regions in which large numbers of rotation periods have
been measured. For the analysis presented here we take rotation
periods from \citealt{herbst02} (ONC), \citealt{lamm04,lamm05} (NGC
2264), \citealt{irwin08} (NGC 2362) and \citealt{littlefair10} (Cep
OB3b). We concentrate on the higher mass stars ($0.4 \le {\rm M} <
1.0$M$_{\odot}$) in these clusters, since these stars show the
clearest difference in rotation rate between stars with discs and
those without \cite[see][for example]{cieza07}, and show little
dependence between rotation rate and stellar mass
\cite[e.g][]{littlefair10}. Masses for stars in each association were
calculated in a consistent manner, using the extinction and
distance-corrected $I$-band magnitudes and the models of
\cite{baraffe98} - see \cite{littlefair10} for more details. For the
ONC, individual extinction values from \cite{hillenbrand97} were used,
but for the other associations a single extinction value was adopted
for all stars. This is unlikely to be a realistic assumption, but
fortunately the reddening vector lies approximately along the
isochrones in a V, V-I diagram and so it is unlikely to affect our
analysis.

\begin{figure}
\begin{center}
$
\begin{array}{cc}
\includegraphics[scale=0.23,angle=0,trim=30 10 30 30,clip]{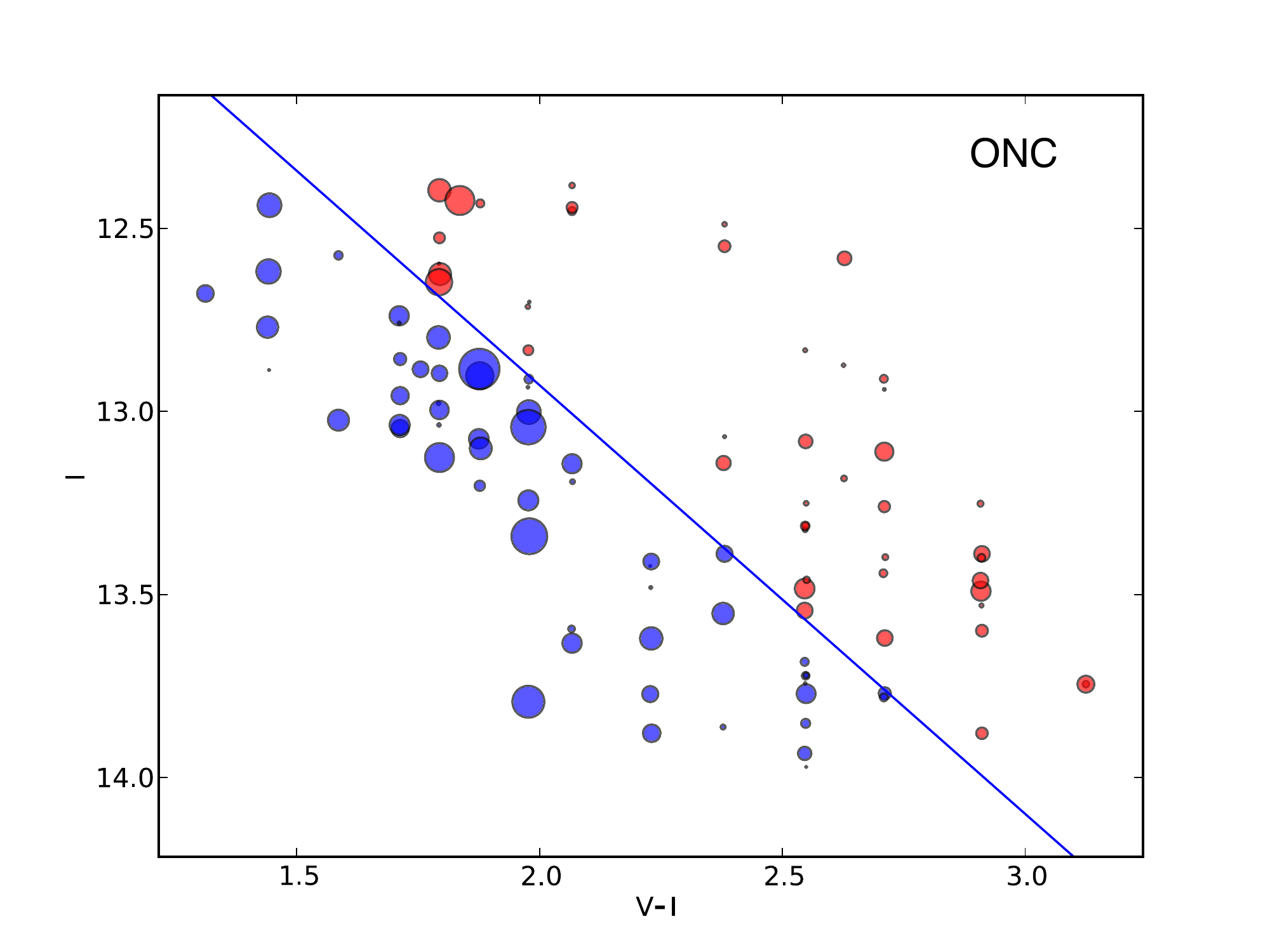} & 
\includegraphics[scale=0.23,angle=0,trim=30 10 30 30,clip]{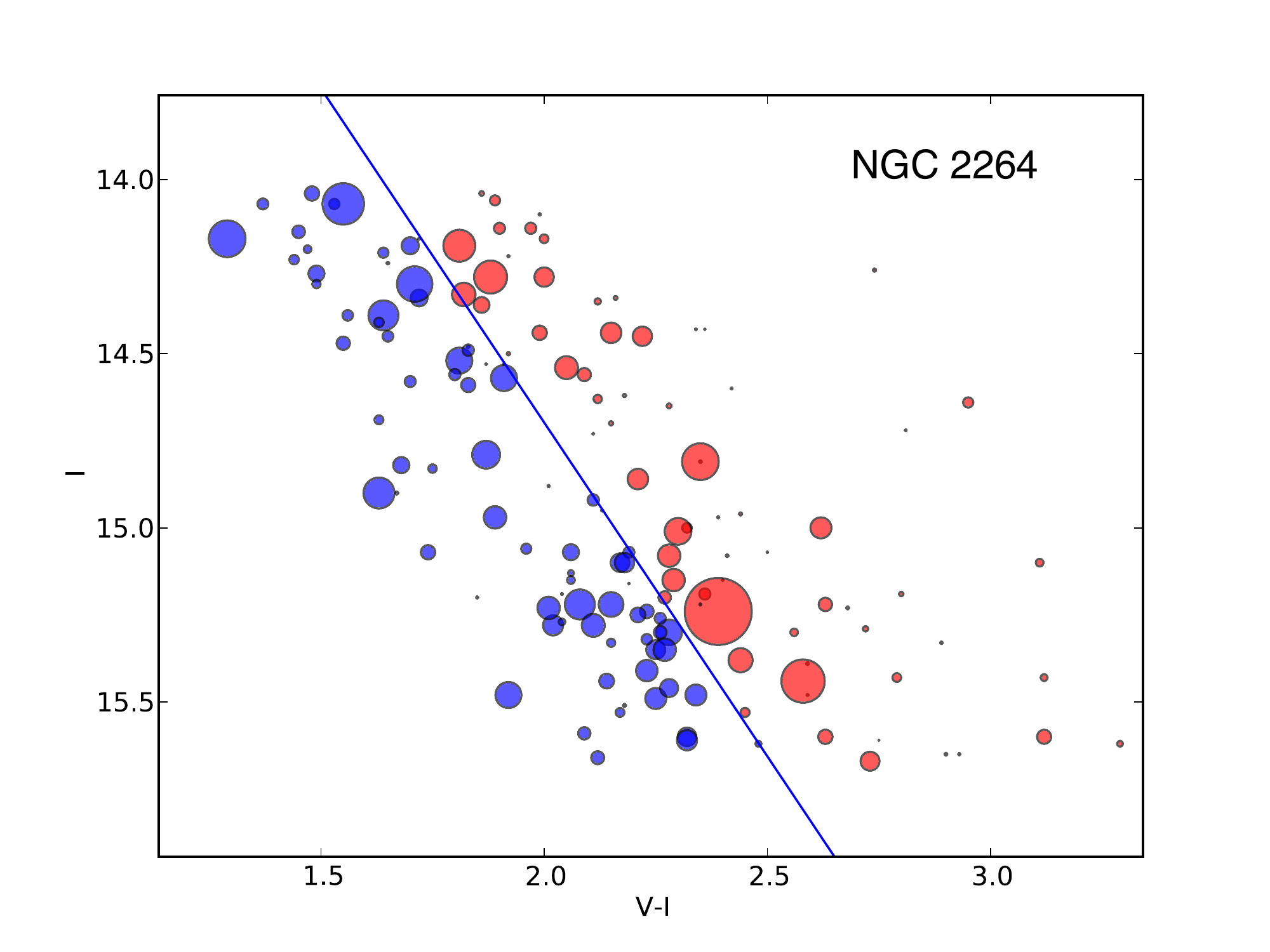} \\ 
\includegraphics[scale=0.23,angle=0,trim=30 10 30 30,clip]{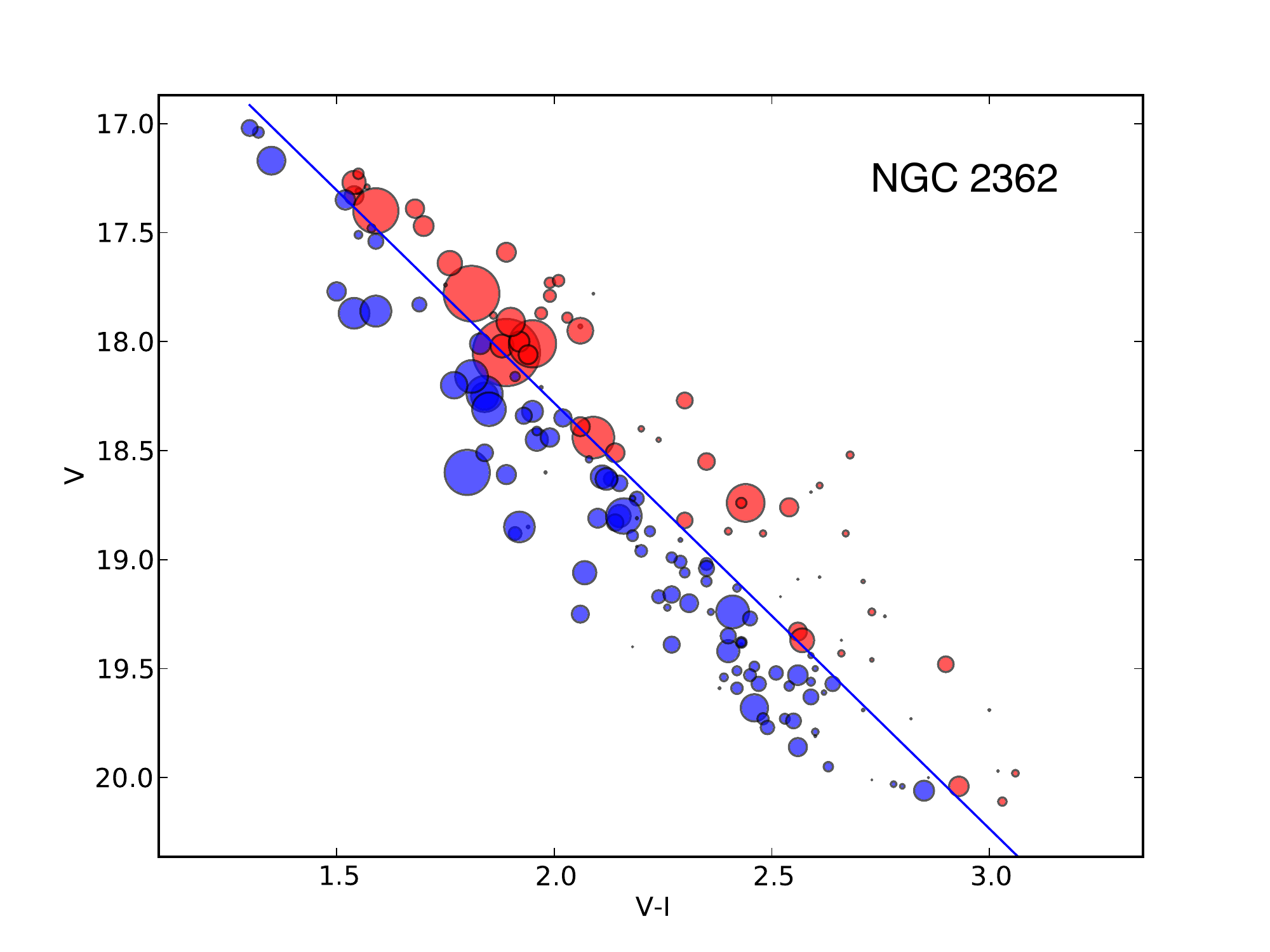} &
\includegraphics[scale=0.23,angle=0,trim=30 10 30 30,clip]{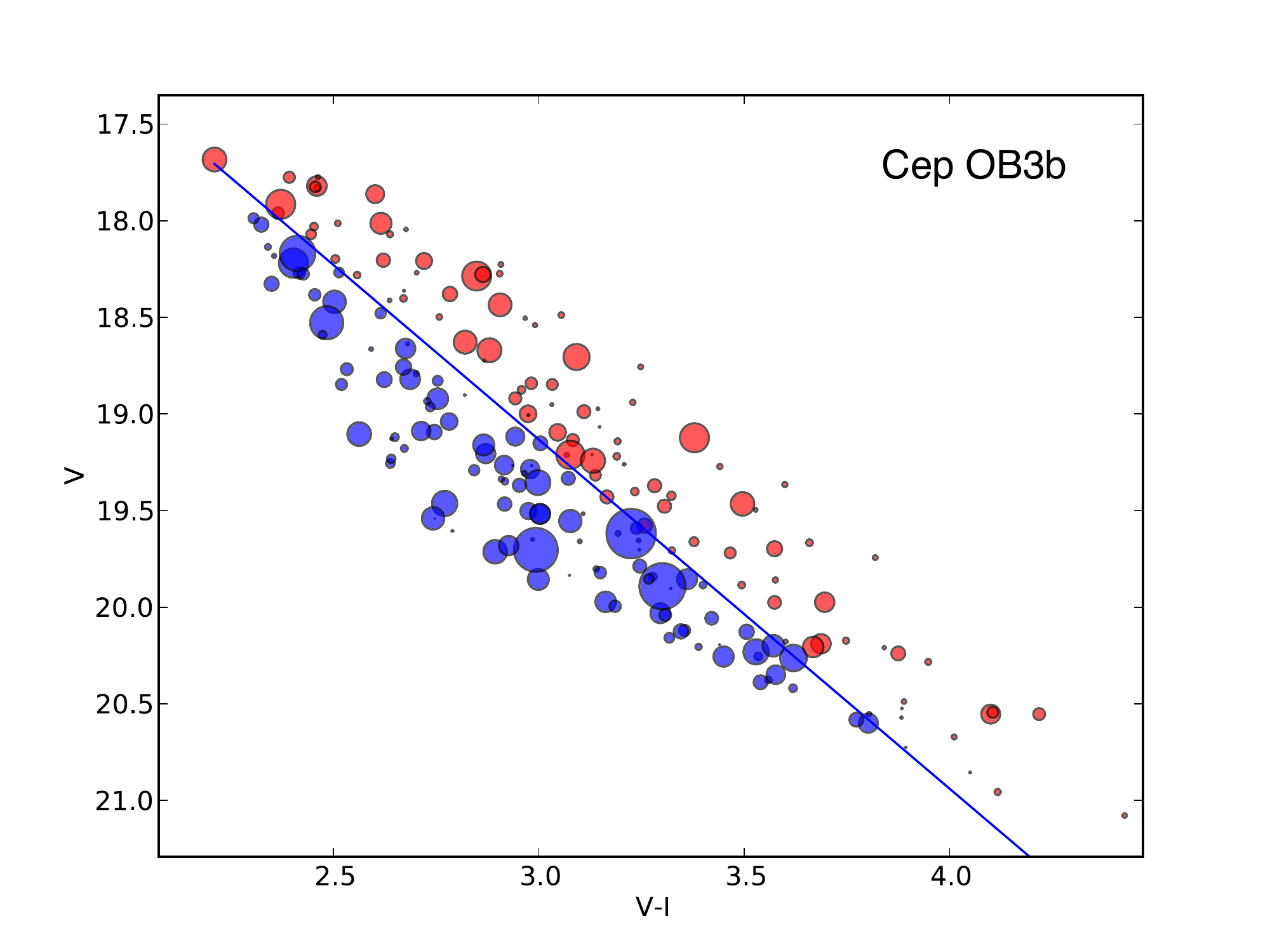} \\
\end{array}
$
\caption{Subdivision of the cluster rotation rate data into bright and faint
  sub-samples. The solid blue line is a linear fit to the median colour in a series 
  of magnitude bins. This fit is used to divide the data into
  a bright sub-sample (red circles) and a faint one (blue circles). 
  The area of a circle is proportional to that star's rotation period}
\label{fig:age_sel}
\end{center}
\end{figure}

We divided our data into bright and faint sub-samples using the
following method. First we calculated the median colour in several
magnitude bins, and a first-order polynomial was fitted to define the
median magnitude as a function of colour. Stars with magnitudes brighter than this were assigned to the bright sub-sample
and vice-versa. The results are shown in figure~\ref{fig:age_sel}.  We note here that the results in this letter are robust against changes in the method described above.
Our results do not change significantly if we fit the median colour in a series of magnitude 
bins, or the median magnitude in a series of colour bins. Nor does the result depend on
the order of the polynomial used to fit the median values. 

\begin{figure}
\begin{center}
\includegraphics[scale=0.42,angle=0,trim=20 20 20 20,clip]{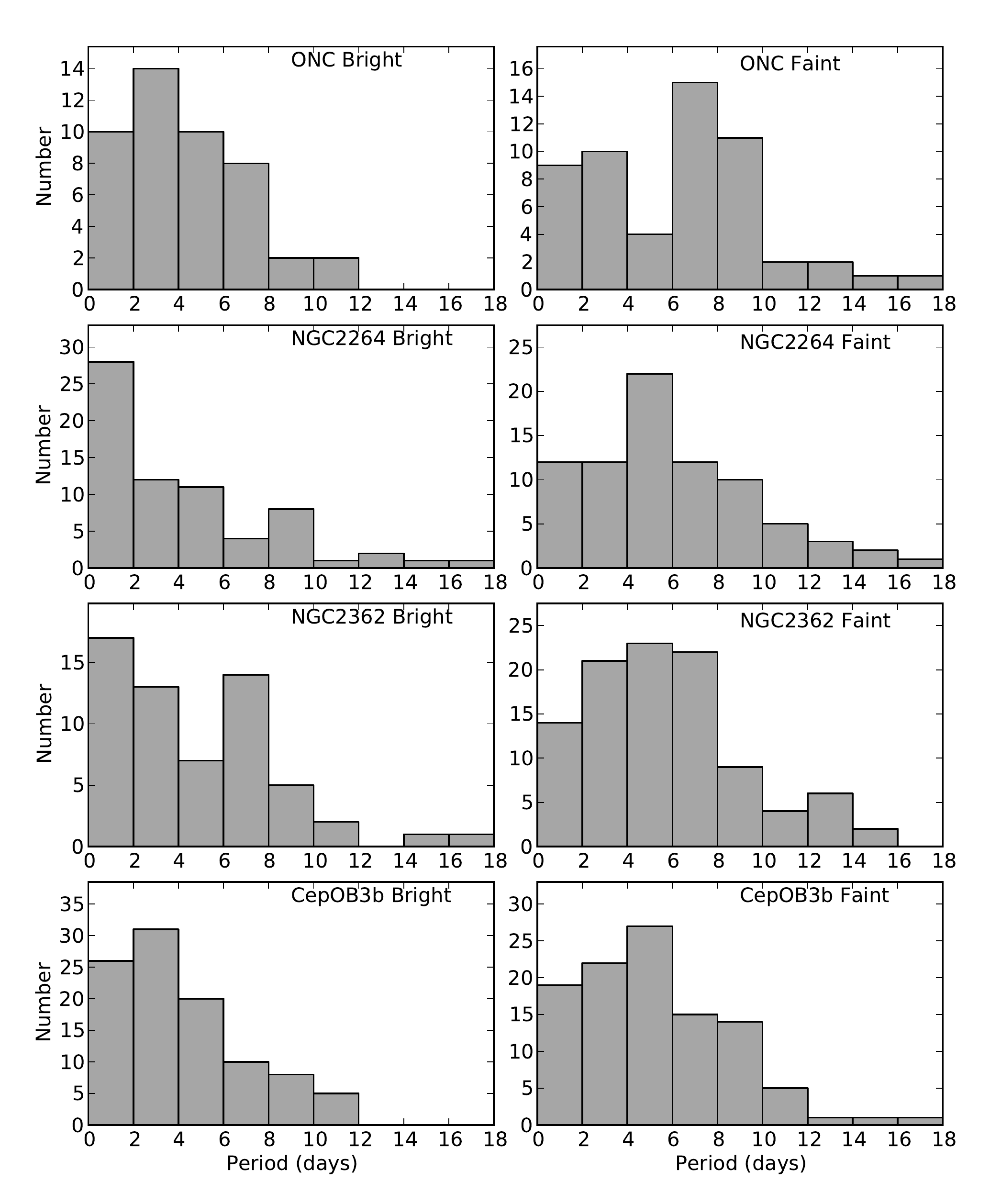} 
\caption{Period distributions for Cep OB3b, NGC 2264, NGC 2362 and the
  ONC, divided into samples according to position in the CMD.  }
\label{fig:pdist_v_age}
\end{center}
\end{figure}

The period distributions for our data are shown in
figure~\ref{fig:pdist_v_age}. It is readily apparent to the eye that,
within each star forming region, the period distributions of the
bright and faint sub-samples differ markedly. For example, in the ONC,
the bright sub-sample lacks the characteristic bi-modal distribution
observed in this cluster, showing instead a rising distribution
towards short periods. By contrast, the faint sub-sample shows the
familiar bi-modal distribution, with a secondary peak of slow rotators
at 6--10 days.

We tested the null-hypothesis that the bright and faint sub-samples
were drawn from the same parent distribution using a 1-D K-S test. The
null hypothesis was rejected with 99.8 percent confidence for the ONC,
99.3 percent confidence for NGC 2264, 99 percent confidence for Cep
OB3b and 98 percent confidence for NGC 2362. In all cases, the
median period of the faint sub-sample is larger than the median period
of the bright sub-sample. Whilst the period distributions of bright
and faint sub-samples overlap substantially, we conclude that, on
average, the more luminous stars in an association rotate more rapidly
than the fainter stars.

It is well known that the lowest mass stars in an association rotate
faster than the high mass stars \cite[see][for
examples]{irwin08,littlefair10,herbst02}. Our sample was carefully
chosen to avoid this confounding factor; the link between rotation and
mass is weak or absent for masses greater than
$\sim$0.4\,M$_{\odot}$. Furthermore, our bright and faint sub-samples
have very similar mass distributions, a 1-D K-S test gives no evidence
that the distributions are different, and the two samples show
identical median masses. Nevertheless, we tested to see if a residual
mass effect could be causing the correlation observed in
figure~\ref{fig:pdist_v_age}. We divided the data from each cluster
into higher mass and lower mass sub-samples, by splitting the data
around the median mass. In all cases, a 1-D K-S test provided no
evidence that the high-mass and low-mass sub-samples were drawn from
different parent distributions. We can therefore rule out the
mass-dependence of rotation as the cause of the correlation between
luminosity and rotation rate.

\begin{figure}
\begin{center}
\includegraphics[scale=0.35,angle=0,trim=50 20 60 0,clip]{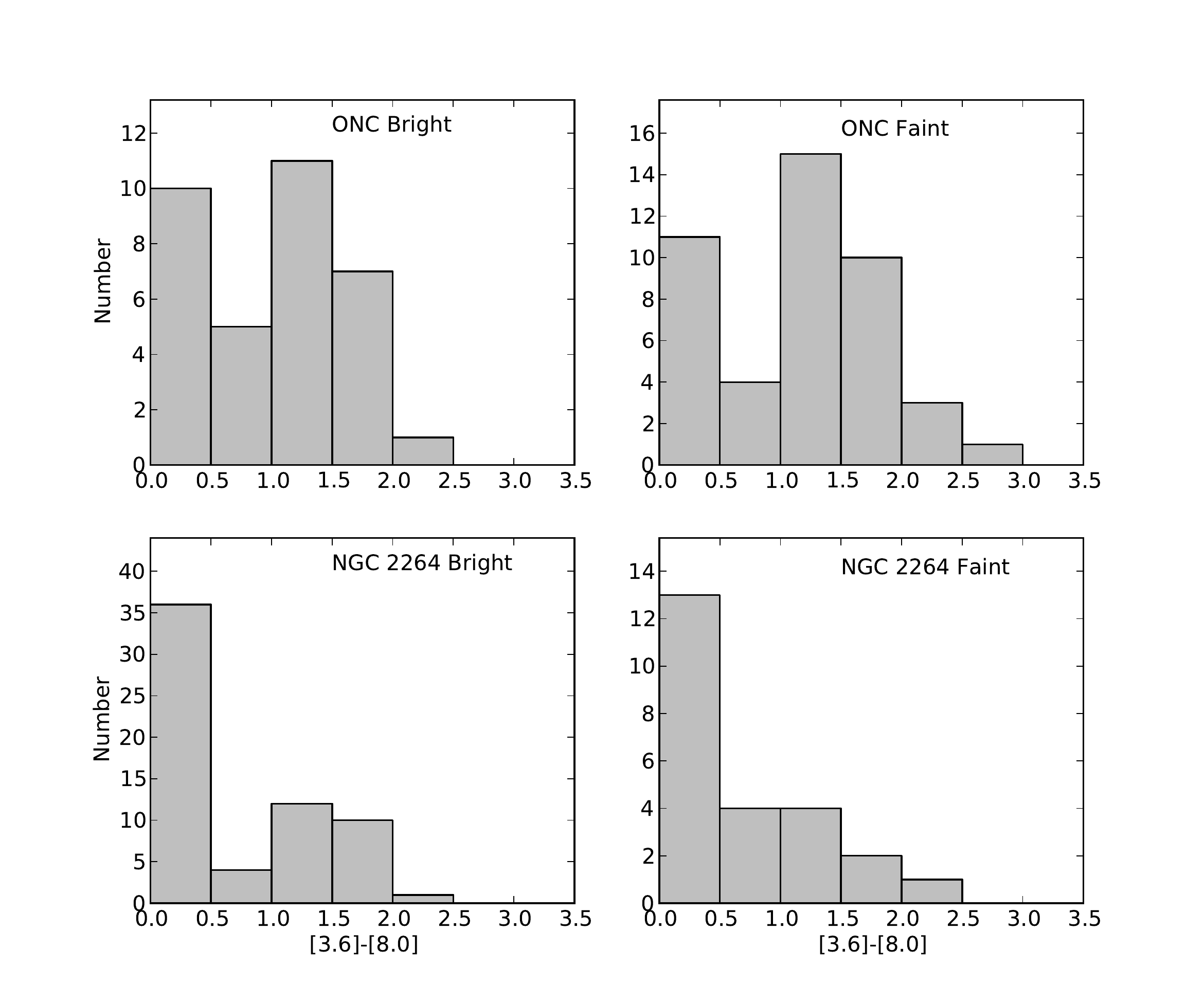} 
\caption{Distributions of [3.6]-8.0] colours for NGC 226 and the ONC,
  divided into samples according to position in the CMD.  }
\label{fig:spitz_v_age}
\end{center}
\end{figure}

It is also possible that luminosity is correlated with the absence or
presence of ongoing accretion. Since rotation rate is itself strongly
correlated with the presence or absence of accretion discs, we might
be observing a secondary correlation arising as a result. To check
this we compared the distribution of Spitzer [3.6]-[8.0] colours
\cite[taken from][]{cieza07} for our bright and faint sub-samples in
those clusters with sufficient existing Spitzer data (ONC and NGC
2264, see Figure~\ref{fig:spitz_v_age}). We found that the fraction of
sources displaying an infrared excess was indistinguishable between
the bright and faint sub-samples. This was confirmed with a 1d K-S
test, which showed no evidence that the distribution of [3.6]-[8.0]
colours in the bright and faint sub-samples were drawn from different
parent distributions.

Lastly, we consider the possibility that the bright and faint
sub-samples are affected differently by contamination from field star
populations. For NGC 2264 and Cep OB3b, multiple colour cuts were used
to remove background objects from the samples, so we believe this is
unlikely in the case of those two associations. Since the correlation
we report is common to all associations, we therefore rule out
contamination by field stars as an explanation.
We conclude that a correlation between rotation rate and luminosity is
a genuine and widespread property of young stellar associations.

\section{Discussion}
\label{sec:discussion} 
If we assume that luminosity spreads in a star forming region
correspond to genuine age spreads, the results presented in
section~\ref{sec:results} are not compatible with the ``disc locking"
model for the spin evolution of young stars. Since accretion discs
disperse over time, the youngest and hence brightest stars are more
likely to possess discs and should rotate more slowly. This is
compounded by the fact that young stars will be larger and will have
larger moments of inertia. Therefore, if age spreads in star-forming
environments are real, we expect the older (and less luminous) stars
to be rotating more rapidly than the younger (and brighter)
stars. This is the opposite trend to that observed. Therefore, either
the luminosity spreads in star forming regions do not primarily
reflect differences in stellar ages, or our understanding of the spin
evolution of young stars needs revising.

The observed correlation between luminosity and rotation is more
easily explained if we start from the assumption of a co-eval
population within which luminosity spreads are introduced due to
differing accretion histories. Starting from this assumption, there
are three plausible mechanisms that can explain our results, which we
refer to as scenarios A, B and C respectively.

\subsection{Scenario A}
In the models of \cite{baraffe09}, stars accrete large fractions of
their final mass during the assembly phase which lasts from
$\sim$10,000 to 100,000 years depending on whether accretion is
episodic or not, although estimates of the lifetimes of the
evolutionary stages of young stars strongly favour episodic accretion
lasting for $\sim100,000$ years \cite[see][and references
therein]{baraffe09}. The heavy accretion rates in the assembly phase
drive stars away from thermal equilibrium. Luminosity spreads are
introduced because stars accrete different fractions of their final
mass during this phase; stars which accrete a large fraction of their
final mass are further from thermal equilibrium, and are much smaller
than stars which accrete a small fraction of their final mass during
this phase.

We assume that all stars emerge from the assembly phase locked to
their discs, and at similar rotation rates, and consider the ensuing
spin-up towards the main sequence once disc-locking ceases. Stars
which rapidly accrete large fractions of their mass during the
assembly phase are already quite small, and undergo limited
contraction and spin up as they approach the main sequence. A star
which accretes a smaller fraction of its final mass during the
assembly phase undergoes much more contraction, and hence spins up
significantly as it approaches the main sequence. Case A thus
naturally produces small, faint stars which are rotating more slowly
than their larger, brighter counterparts.

\subsection{Scenario B}
In this scenario, we assume that the stars which accrete large
fractions of their final mass during the assembly phase are likely to
be those in dense environments surrounded by massive discs. We further
assume that these stars are more likely to emerge from the assembly
phase surrounded by massive discs which will in turn be longer-lived
and supply the pre-main-sequence star with material at a higher
accretion rate. These stars are most likely to remain locked to their
discs and will show slow rotation rates. Since they experienced heavy
assembly phases, they are also the smallest, faintest stars.

By contrast, stars which are rotating rapidly at a few Myr must, according to the disc-locking model, have been released from their discs after $\sim10^5$\, yr. The most rapidly rotating stars in an
association are therefore those stars with shortest disc
lifetimes. Under scenario B, these stars also accreted a small
fraction of their final mass during the assembly phase and are
therefore the largest, brightest stars.

\subsection{Scenario C}
Some authors have suggested \cite[e.g.][]{hartmann96,kley99} that, contrary to the results of \cite{baraffe09}
phases of rapid accretion can dramatically {\em increase} the size (and luminosity) of a young star.
Such a star would look much younger in a CMD than a non-accreting counterpart with the same
initial mass and radius. The accreting star's Kelvin-Helmholtz timescale drops as a result
of the increased size and luminosity. Thus, over a given time period the accreting star contracts
and spins up more than it's non-accreting counterpart. Thus, some time after the accretion event, the accreting star looks younger, and will be rotating more rapidly than, the non-accreting star. In other words, such a scenario can also explain the correlations presented here.


In summary, there are three plausible scenarios which explain the
correlation between luminosity and rotation. Each starts from a co-eval
population into which luminosity spreads are injected through varying
accretion histories. We therefore argue that the observed correlation
between luminosity and rotation is strong evidence that luminosity
spreads in star forming region are primarily caused by a range of
accretion histories, instead of a spread in ages.

In principle, it ought to be possible to determine if
accretion-induced luminosity spreads can explain the full extent of
the observed luminosity spreads in star forming regions by
simultaneously modelling the affects of accretion on the stellar
luminosity and spin rate. The difficulty with such an analysis at
present lies in the fact we do not know how accretion at a given rate
effects the spin of the central object. Proper modelling of this
result will therefore require a fuller understanding of the physical
mechanism(s) by which the star-disc interaction governs stellar
rotation. In addition, some of the scatter seen in
figure~\ref{fig:age_sel} results from variability, extinction and
binarity. This scatter might also explain some of the substantial
overlap between the period distribution of bright and faint
samples. Any analysis would have to include a careful treatment of
these confounding factors.

\subsection{Alternatives}
Can any other mechanisms plausibly explain the sense of the
correlation we observe here? One potential mechanism is the effect of
magnetic activity on stellar radius. Observations of low-mass stars in
eclipsing binaries show that they are over-sized by some 5--10 percent
\cite[see][for a review]{ribas06}. Several authors have argued that
this could be the result of magnetic activity \citep{lopez-morales07,
  ribas08}, and \cite{chabrier07} showed that this discrepancy could
be explained by magnetic effects, including starspots. Since
pre-main-sequence stars are magnetically highly active
\citep[e.g.][]{donati10} it is plausible this mechanism operates in
pre-main-sequence stars too. We can speculate that the influence of
magnetic fields on stellar structure might cause a correlation similar
in {\em sense} to that presented in this paper. The effect of starspot
coverage is to increase the stellar radius and reduce the effective
temperature, at roughly constant luminosity \citep{chabrier07}. If the
lower effective temperature produces redder V-I colour, and if rapid
rotators have larger spot coverage than slower rotators, this could
explain the correlation observed here.  However, neither of these
conditions are likely to be satisfied. So far, there is little
evidence for a link between rotation and activity for
pre-main-sequence stars \cite{scholz07}; it is likely that most
pre-main-sequence stars show saturated levels of activity across a
wide range of rotation rates. Even if rapid rotators did have higher
spot coverages, it is far from clear that the ensuing drop in
effective temperature would cause a change in V-I colour. At the
typical effective temperatures within our sample (T$_{\rm eff} \sim
3500$\,K), a cool starspot reduces both the V-band and I-band flux,
and can leave V-I unchanged. Good evidence that this in fact the case
comes from the CMD of intermediate age clusters like NGC 2547
\citep[e.g.][]{jeffries04}, which have a very tightly defined
pre-main-sequence locus in a V, V-I CMD, and contain many stars with
saturated levels of magnetic activity \citep{jeffries06}. Therefore, we
believe it is unlikely that starspots are the origin of the
correlation presented within this paper.

\section{Conclusions}
\label{sec:concl}
We have presented an analysis of the rotation of young stars in the
associations Cepheus OB3b, NGC 2264, NGC 2362 and the Orion Nebula
Cluster (ONC). The rotation rate shows a significant correlation with
position in a CMD, in the sense that stars with below average
luminosity are rotating more slowly than stars with above average
luminosity. If position within the CMD is interpreted as being due to
genuine age spreads within a cluster, then the implication is that the
youngest stars in the cluster (those with the largest moments of
inertia and highest likelihood of ongoing accretion) are the most
rapidly rotating. Such a result is in conflict with the existing
picture of angular momentum evolution in young stars, where the stars
are braked effectively by their accretion discs until the disc
disperses.

Instead, we argue that, for a given association of young stars,
position within the CMD is not primarily a function of age, but of
accretion history. We have shown that this hypothesis can, in
principle, explain the correlation we observe between rotation rate
and position within the CMD. Since variations in accretion history can lead to spreads 
in radii and luminosity matching those observed, evidence for age spreads in star forming
regions must currently be viewed sceptically. Furthermore, this
implies that masses and ages, which are inferred by comparison to
non-accreting evolutionary tracks, may be in error. If true, the
initial mass functions and ages of star forming regions may need
significant revision.
\section*{\sc Acknowledgements}

SPL is supported by an RCUK fellowship.  This research has made use of
NASA's Astrophysics Data System Bibliographic Services.

\bibliographystyle{mn2e}
\bibliography{abbrev,refs,refs2}

\end{document}